\documentclass[%
reprint,
superscriptaddress,
amsmath,
amssymb,
aps,
prx,
floatfix,
]{revtex4-2}
\usepackage{graphicx}
\usepackage{dcolumn}
\usepackage{bm}
\usepackage{hyperref}
\usepackage{xr}
\usepackage{lipsum}
\usepackage{braket}
\usepackage{booktabs}
\usepackage{xcolor}
\preprint{APS/123-QED}
\begin{document}

\title[Photon-mediated long range coupling of two Andreev level qubits]{Photon-mediated long range coupling of two Andreev level qubits}

\author{L.\,Y.~Cheung}
\affiliation{Department of Physics, University of Basel, Klingelbergstrasse 82, CH-4056, Switzerland}

\author{R.~Haller}
\affiliation{Department of Physics, University of Basel, Klingelbergstrasse 82, CH-4056, Switzerland}

\author{A.~Kononov}
\affiliation{Department of Physics, University of Basel, Klingelbergstrasse 82, CH-4056, Switzerland}

\author{C.~Ciaccia}
\affiliation{Department of Physics, University of Basel, Klingelbergstrasse 82, CH-4056, Switzerland}

\author{J.\,H.~Ungerer}
\affiliation{Department of Physics, University of Basel, Klingelbergstrasse 82, CH-4056, Switzerland}
\affiliation{Swiss Nanoscience Institute, University of Basel, Klingelbergstrasse 82, CH-4056, Switzerland}

\author{T.~Kanne}
\affiliation{Center for Quantum Devices, Niels Bohr Institute, University of Copenhagen, DNK-2100, Denmark}

\author{J.~Nygård}
\affiliation{Center for Quantum Devices, Niels Bohr Institute, University of Copenhagen, DNK-2100, Denmark}

\author{P.~Winkel}
\affiliation{IQMT, Karlsruhe Institute of Technology, DE-76344, Germany}
\affiliation{Departments of Applied Physics and Physics, Yale University, Connecticut 06511, USA}
\affiliation{Yale Quantum Institute, Yale University, Connecticut 06520, USA}

\author{T.~Reisinger}
\affiliation{IQMT, Karlsruhe Institute of Technology, DE-76344, Germany}

\author{I.\,M.~Pop}
\affiliation{IQMT, Karlsruhe Institute of Technology, DE-76344, Germany}

\author{A.~Baumgartner}
\affiliation{Department of Physics, University of Basel, Klingelbergstrasse 82, CH-4056, Switzerland}
\affiliation{Swiss Nanoscience Institute, University of Basel, Klingelbergstrasse 82, CH-4056, Switzerland}

\author{C.~Sch{\"o}nenberger}
\homepage{www.nanoelectronics.unibas.ch}
\affiliation{Department of Physics, University of Basel, Klingelbergstrasse 82, CH-4056, Switzerland}
\affiliation{Swiss Nanoscience Institute, University of Basel, Klingelbergstrasse 82, CH-4056, Switzerland}

\date{\today}

\begin{abstract}
In a superconducting weak link, the supercurrent is carried by Andreev bound states (ABSs) formed by the phase-coherent reflection of electrons and their time-reversed partners. A single, highly transmissive ABS can serve as an ideal, compact two-level system, due to a potentially large energy difference to the next ABS \cite{PhysRevLett.90.087003}. While the coherent manipulation of such Andreev levels qubits (ALQs) has been demonstrated \cite{PhysRevLett.121.047001,doi:10.1126/science.aab2179}, a long-range coupling between two ALQs, necessary for advanced qubit architectures \cite{doi:10.1038/nature06184,Borjans2020}, has not been achieved, yet. Here, we demonstrate a coherent remote coupling between two ALQs, mediated by a microwave photon in a novel superconducting microwave cavity coupler. The latter hosts two modes with different coupling rates to an external port. This allows us to perform fast readout of each qubit using the strongly coupled mode, while the weakly coupled mode is utilized to mediate the coupling between the qubits. When both qubits are tuned into resonance with the latter mode, we find excitation spectra with avoided-crossings, in very good agreement with the Tavis-Cummings model \cite{PhysRev.170.379}. Based on this model, we identify highly entangled two-qubit states for which the entanglement is mediated over a distance of six millimeters. This work establishes ALQs as compact and scalable solid-state qubits.
\end{abstract}

\maketitle

\begin{figure*}[tb]
	\centering
	\includegraphics[width=\linewidth]{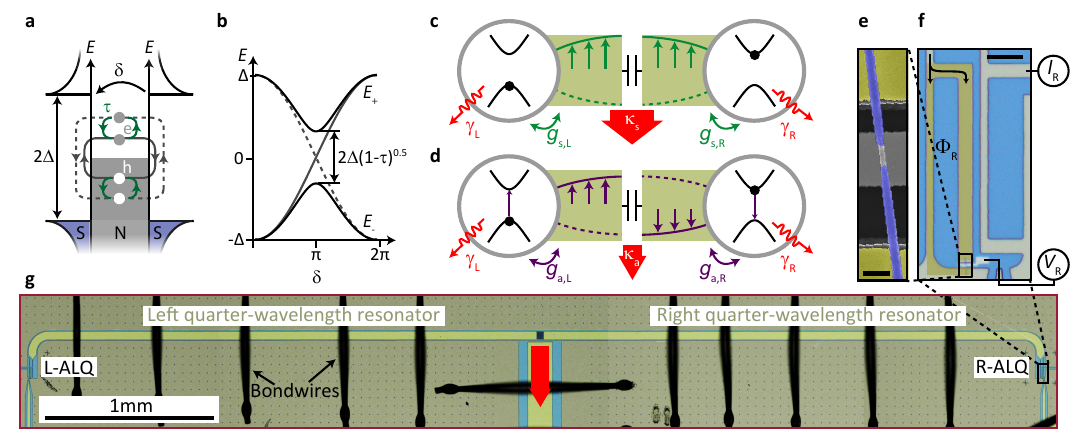}
    \caption{\textbf{Andreev level qubit (ALQ) coupling device.} \textbf{a,} Andreev bound state (ABS) formation in an S-N-S weak link. In the N region, electron (e) and hole (h) partial waves can constructively interfere to form discrete subgap levels below the gap $\Delta$. The phases of the partial waves obtain contributions from Andreev reflections at the N-S interfaces (curved grey lines) and from the propagation in the N-region (horizontal grey lines). The two spin-degenerate and time-reversed trajectories can be coupled by single-particle scattering (green lines) with probability $\tau$. \textbf{b,} The discrete Andreev levels $E_\mathrm{\pm}(\delta)$ are plotted as a function of the phase difference $\delta$ between the two superconducting reservoirs. An energy gap of $2\Delta\sqrt{1-\tau}$ is opened by a finite $\tau$ at $\delta = \pi$. \textbf{c-d,} Schematics of the complete device containing two ALQs (circles) with qubit decay rates $\gamma_\mathrm{L/R}$ and ALQ-coupler coupling strength $g_\mathrm{s/a,L/R}$ to the two cavity coupler modes. The symmetric mode (green) is used to read out qubit states, while the anti-symmetric mode (purple) is used to coupling the two ALQs, with mode dependent coupling rate $\kappa_\mathrm{a}\ll\kappa_\mathrm{s}$ to the readout port. \textbf{e,} A scanning electron micrograph of the right ALQ (R-ALQ) with epitaxially grown superconducting Al shells (purple), suspended over a metallic bottom gate (grey). The scale bar is 300\,nm. \textbf{f,} An optical micrograph of the qubit control lines. The junction is galvanically connected to the center conductor in a RF-SQUID geometry (yellow), such that the qubit-coupler coupling is mediated by the currents in the common superconducting leads. The qubit frequency is controlled by a DC gate voltage $V_\mathrm{R}$ and the phase control $\delta_\mathrm{R}$ is generated by a DC current $I_\mathrm{R}$ in a flux line. The scale bar is 10\,$\mathrm{\mu m}$. \textbf{g,} Composite optical micrographs of the full device. It consists of two capacitively coupled quarter-wavelength coplanar transmission line resonators (light green).} 
	\label{fig:fig1}
\end{figure*}

The fundamental quantum states in superconducting weak links are Andreev bound states (ABSs) that form as superpositions of propagating electrons and holes near a superconductor \cite{osti_4071988,kulik,jünger2021intermediate}. In Fig.\,\ref{fig:fig1}a), we illustrate the formation of a single, highly transmissive ABS in a short normal metal or semiconductor region (N) between two superconducting reservoirs (S) with a superconducting gap $\Delta$. The condition for constructive interference of the electron and hole partial waves contains phase shifts due to Andreev reflections at the NS interfaces and the propagation in N \cite{Jünger2019,PhysRevB.46.12573,kulik1969macroscopic}. For a single channel in the short junction limit, the constructive interference results in two time-reversed, spin-degenerate Andreev levels, with eigenenergies tuned by the phase difference $\delta$ between the two superconducting order parameters. Including a transmission probability $\tau$ in the N part between the two superconductors, one obtains the energy spectrum $E_\pm(\delta)=\pm\Delta \sqrt{1-\tau\sin^2(\delta/2)}$ around the Fermi energy, as shown in Fig.\,\ref{fig:fig1}b) \cite{PhysRevLett.90.087003,PhysRevB.43.10164,PhysRevLett.66.3056,FURUSAKI1999809}. We choose these two states at constant $\delta=\pi$ to define the Andreev level qubit (ALQ) with a tunable excitation gap of $\Delta E=2\Delta \sqrt{1-\tau}$. The corresponding qubit transition frequency is $f_\mathrm{qb}=\Delta E/h$, with the Planck constant $h$. Since the next ABSs are typically found at much larger energies near $\Delta$, the next excited state can in principle be engineered with a much greater energy difference than the qubit transition. This unique gate-tunable energy spectrum stands in strong contrast to other, more established superconducting qubits, e.g., transmon qubits \cite{PhysRevA.76.042319}, in which dynamical driving of the qubits is severely limited by the leakage out of the computational subspace into higher excited states \cite{McEwen2021,Werninghaus2021}.

To implement a general quantum algorithm, it is necessary to couple two qubits coherently. While the coupling between two ABSs over short distances is being explored in various systems \cite{jünger2021intermediate,Su2017,Kürtössy2021,pitavidal2023strong}, a long-distance coupling could not be established so far. On the other hand, the versatile long-distance coupling and quantum state readout have been established for other qubit platforms, for example superconducting \cite{doi:10.1038/nature06184} or semiconductor qubits \cite{Borjans2020,PhysRevX.8.041018}, using superconducting microwave resonators. The reproducibility and low losses of superconducting microwave resonators, and the potentially strong coupling to the ALQs make these techniques ideal to transfer quantum information between ALQs \cite{PhysRevResearch.3.013036}.

Despite significant progress in understanding coupled ABS-resonator systems using circuit quantum electrodynamics, only experiments with single ALQs have been performed so far \cite{PhysRevResearch.3.013036,doi:10.1126/science.aab2179,PhysRevLett.121.047001,PhysRevX.9.011010,PhysRevLett.128.197702,https://doi.org/10.48550/arxiv.2208.10094,doi:10.1126/science.abf0345,doi:10.1038/s41567-020-0952-3,PhysRevB.96.125416,PhysRevResearch.3.L022012,PhysRevB.104.174517,PhysRevB.96.125416}. Here, we first demonstrate a strong coupling of two individual ALQs to the same superconducting resonator mode, with qubit decay rates lower than the coupling strengths to the resonator mode. In a second step, we simultaneously bring the ALQ transition energies into resonance with a specific resonator mode that only couples weakly to the measurement circuit. A coherent, exchange-type coupling between the two ALQs is then established, with the mutual coupling mediated by a microwave photon, and not disturbed by the out-coupling. The resulting excitation spectrum is fully captured by the Tavis-Cummings model, solely using independently determined single qubit parameters \cite{PhysRev.170.379}. This allows us to identify highly entangled two-qubit states, possibly allowing the implementation of future remote two-qubit gate operations for Andreev qubits.

The investigated quantum circuit is illustrated schematically in Fig.\,\ref{fig:fig1}c-d) and consists of two InAs nanowire weak links, each with a single, highly transparent ABS forming the ALQ. The nanowires have an epitaxially grown Al shell \cite{Krogstrup2015}, etched away in a short semiconducting region to form the left ALQ (L-ALQ) and the right ALQ (R-ALQ). Local bottom gates are used to separately control the qubit frequency of each ALQ. A scanning electron micrograph of the R-ALQ is shown in  Fig.\,\ref{fig:fig1}e). This nanowire weak link is embedded in a superconducting pick-up loop forming an rf-SQUID \cite{tinkham2004introduction}, as shown in Fig.\,\ref{fig:fig1}f). The threading magnetic fluxes $\Phi_{\rm L,R}=\Phi_0\delta_\mathrm{L,R}/2\pi$, controlled by an external current line, is used to set the corresponding phase differences $\delta_\mathrm{L,R}$, with $\Phi_0 = h/2e$ the superconducting flux quantum \cite{PhysRevResearch.3.013036}. The cavity coupler is composed of two nominally identical quarter-wavelength resonators (Fig.\,\ref{fig:fig1}g). One of the key elements in our design is a coupling capacitor designed between the two resonators at the respective voltage anti-nodes. The coupling capacitor prohibits low frequency dissipative current flowing through the center conductor and thus ensures low internal loss of the cavity coupler. This design combines the two quarter-wavelength modes of the individual coplanar transmission line resonators into two half-wavelength modes, one mode with a symmetric voltage profile along the transmission line with a resonance frequency of $f_\mathrm{s} = 6.461\,\mathrm{GHz}$ (Fig.\,\ref{fig:fig1}c) and the other mode with an anti-symmetric voltage profile with a resonance frequency of $f_\mathrm{a} = 6.075\,\mathrm{GHz}$ (Fig.\,\ref{fig:fig1}d). The readout port, designed as an open-ended coplanar transmission line to the center of the cavity coupler, can be used to apply a symmetric voltage excitation to the cavity coupler and thus couples strongly to the symmetric coupler mode with a coupling strength $\kappa_\mathrm{s}/2\pi = 1.2\,\mathrm{MHz}$. In contrast, the excitation of the anti-symmetric coupler mode through the readout port is suppressed due to the mismatch of the symmetries between the voltage excitation and coupler mode. The pick-up loops are galvanically connected to the central conductor of the respective resonator current anti-nodes (Fig.\,\ref{fig:fig1}f). 

We first demonstrate a strong coupling between each ALQ and the anti-symmetric coupler mode. Owing to a weak coupling to the readout port, the anti-symmetric coupler mode has a long lifetime and can strongly hybridise with an ALQ at small detunings ($f_\mathrm{qb}\approx f_\mathrm{a}$). We probe the excitation spectra of the qubit-coupler hybrid system using pulsed two-tone spectroscopy techniques. The transition from the ground state of the qubit-coupler system to an excited state can be addressed by an excitation pulse with variable microwave frequencies $f_\mathrm{drive}$. The population of such a hybrid excited state shifts the resonance frequency of the symmetric coupler mode, measured in the reflection coefficient of a microwave probe pulse at a frequency near $f_\mathrm{s}$, routed through a Josephson parametric amplifier operating close to the quantum limit \cite{PhysRevApplied.13.024015}. The ALQ-coupler coupling strength for the symmetric coupler mode $g_\mathrm{s,L/R}/(2\pi)$ is measured to be $\sim 120\,\mathrm{MHz}$ (not shown). In our experiment, both the qubit excitation and resonator probe pulses are routed through the readout port to the qubit-coupler system. The resonator probe pulse is applied twice, once during the qubit excitation and the other time after the system relaxes back to the ground state (Details in the Methods). The differential quadrature $ \Delta Q$ is obtained by subtracting the two reflected probe pulses to reject slow drifts. 

\begin{figure*}[tb]
	\centering
	\includegraphics[width=\linewidth]{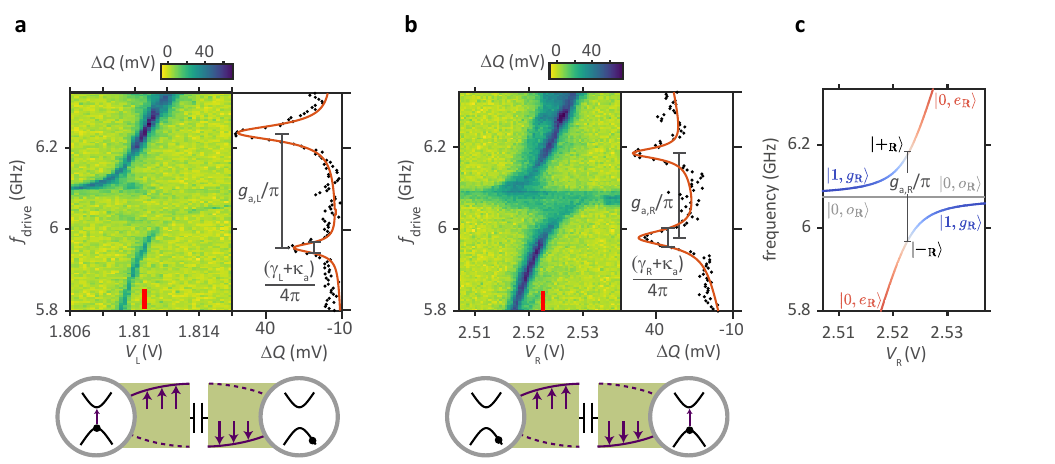}
	\caption{\textbf{Strong coupling between individual ALQs and the anti-symmetric coupler mode.} \textbf{a,} The differential quadrature $\Delta Q$ between the two reflected resonator probe pulses at $f_\mathrm{s}$ as a function of the qubit excitation frequency $f_\text{drive}$ and $V_\text{L}$ at $\delta_\mathrm{L}=\pi$, revealing an avoided crossing between the qubit and coupler resonances. The qubit-resonator coupling rate of $g_\mathrm{c,L}/(2\pi) = 141\,\mathrm{MHz}$ and the qubit decay rate of $\gamma_\mathrm{L}/(2\pi) = 77\,\mathrm{MHz}$ can be directly read out at $V_\mathrm{L}=1.810\,\mathrm{V}$, as illustrated in the cross-section on the right of the plot. An illustration of the interaction between the resonator and the L-ALQ is shown in the bottom insert. The interaction with the R-ALQ is suppressed by tuning the R-ALQ frequency to a larger value with $\delta_\mathrm{R}=1.2\,\pi$, while keeping $V_\mathrm{R}=2.5225\,\mathrm{V}$. \textbf{b,} Similarly, an individual strong coupling between the R-ALQ and the anti-symmetric coupler mode is observed at $V_\mathrm{R}=2.5225\,\mathrm{V}$, with $g_\mathrm{c,R}/(2\pi) = 104\,\mathrm{MHz}$ and $\gamma_\mathrm{R}/(2\pi) = 82\,\mathrm{MHz}$. The horizontal resonance at $f_\mathrm{drive}\approx 6.075\,\mathrm{MHz}$ occurs as the occupation of the odd state due to quasiparticle poisoning. It is missing in \textbf{a} because of small excitation power seen by the circuit. \textbf{c,} Calculated excitation spectrum as a function of the $V_\mathrm{R}$ using the listed parameters, quantitatively reproducing the measured spectrum in \textbf{b}). The two resonances correspond to the qubit-coupler hybrid states $\vert \pm_\mathrm{R}\rangle$, with the color indicating the single qubit weight $\vert \langle 0,e_\mathrm{R} \vert \pm_\mathrm{R} \rangle \vert ^2$.}
	\label{fig:fig2}
\end{figure*}

To investigate a single ALQ-coupler system, we tune the respective phase of the investigated ALQ to the sweet-spot, $\delta_\mathrm{i} = \pi$, where the qubit frequency is insensitive to small phase fluctuations. The other, idling ALQ is phase biased such that its transition frequency is far above $f_\mathrm{a}$. Fig.\,\ref{fig:fig2}a) shows the differential quadrature $\Delta Q$ as a function of $f_\text{drive}$ and $V_\text{L}$. We find an avoided crossing between the qubit and cavity resonances, demonstrating a coherent coupling between the L-ALQ and the anti-symmetric coupler mode \cite{RevModPhys.93.025005}. A fit of $\Delta Q$ with two Lorentzians and a linear background on resonance, $f_\mathrm{a}=f_\mathrm{qb,L}$, results in a qubit-coupler coupling rate of $g_\mathrm{a,L}/(2\pi) = 141\,\mathrm{MHz}$, which exceeds the qubit decay rate $\gamma_\mathrm{L}/(2\pi) = 77\,\mathrm{MHz}$. We find a similarly strong qubit-resonator coupling for the R-ALQ, with $g_\mathrm{a,R}/(2\pi) = 104\,\mathrm{MHz}$ and $\gamma_\mathrm{R}/(2\pi) = 82\,\mathrm{MHz}$ (Fig.\,\ref{fig:fig2}b).

In Fig.\,\ref{fig:fig2}b), we find a weak, gate-independent resonance at a frequency $f_\mathrm{a}=6.075\,\mathrm{GHz}$ between the ALQ-coupler hybrid states. We attribute this line to the excitation of the bare anti-symmetric coupler mode without a coupled ALQ. In the short junction limit, there are two degenerate current-less odd parity states, that are insensitive to the resonator current and thus do not couple to the resonator \cite{PhysRevLett.121.047001}. Parity switching with a rate faster than the measurement time can thus result in a reflected signal by exciting the bare anti-symmetric resonator mode. 


When a bare qubit state is degenerate with the cavity excitation, they form a new set of eigenstates due to the exchange of excitations between them. This new set of qubit-coupler hybrid states is best seen as an avoided crossing in the spectrum described by the Jaynes-Cummings Hamiltonian \cite{1443594,PhysRevResearch.3.013036,doi:10.1126/science.aab2179,PhysRevX.9.011010,PhysRevLett.128.197702}. The measured spectra are very well reproduced by the latter for one resonator mode coupled to a qubit $H_\mathrm{JC} = \hbar g_\mathrm{a,i} (a\sigma_\mathrm{+,i}+a^\dagger\sigma_\mathrm{-,i})$, with $\sigma_\mathrm{+}$ ($\sigma_\mathrm{-}$) and $a^\dagger$ ($a$) being the qubit and resonator raising (lowering) operators. It describes the coherent exchange of one excitation between a qubit $i$
and the cavity mode at a rate of $g_\mathrm{a,i}$. In Fig.\,\ref{fig:fig2}c), we plot the corresponding calculated spectrum for the R-ALQ as a function of $V_\mathrm{R}$ using the extracted $g_\mathrm{a,R}$ (See Methods for Details). The voltage axis is calibrated by equating the lowest and highest resonance frequencies with the corresponding qubit frequencies. The two resonances corresponding to the qubit-coupler hybrid states $\vert \pm_\mathrm{R}\rangle$ are colored according to the qubit weight $\vert \langle 0,e_\mathrm{R} \vert \pm_\mathrm{R} \rangle \vert ^2$. As the R-ALQ qubit frequency is tuned across $f_\mathrm{a}$ by $V_\mathrm{R}$, the eigenstate $\vert +_\mathrm{R}\rangle$ evolves gradually from a bare photonic state $\vert 1,g_\mathrm{R} \rangle$ to a bare qubit excited state $\vert 0,e_\mathrm{R} \rangle$, while $\vert -_\mathrm{R}\rangle$ evolves from $\vert 0,e_\mathrm{R} \rangle$ to $\vert 1,g_\mathrm{R} \rangle$. At the degeneracy point at $V_\mathrm{R}=2.5225\,\mathrm{V}$, the coherent exchange of one excitation forms the two maximally hybridized states $\vert \pm_\mathrm{R}\rangle = 1/\sqrt{2}(\vert 1,g_\mathrm{R} \rangle \pm \vert 0,e_\mathrm{R} \rangle)$, with a vacuum Rabi splitting of $g_\mathrm{a,R}/\pi$. The odd state $\vert 1,o_\mathrm{R}\rangle$ does not couple to the resonator, so that the resonance is purely given by the bare anti-symmetric mode frequency $f_\mathrm{a}$.

\begin{figure*}
	\centering
	\includegraphics[width=\linewidth]{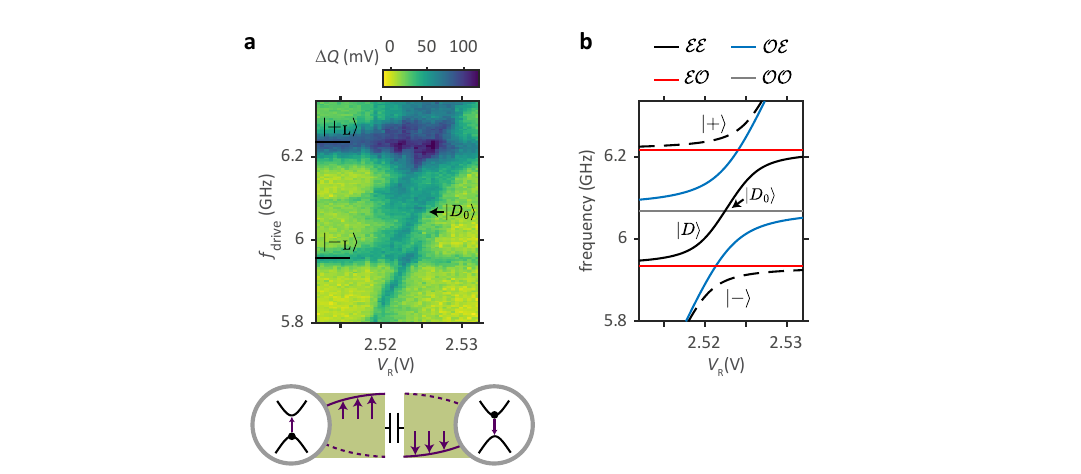}
	\caption{\textbf{Remote coupling of two ALQs.} \textbf{a,} The differential quadrature $\Delta Q$ plotted as a function of $V_\mathrm{R}$ for the case when both ALQs are tuned into resonance with the anti-symmetric resonator mode ($\delta_\mathrm{L}=\delta_\mathrm{R}=\pi,\,V_\mathrm{L}=1.810\,\mathrm{V}$). The states causing the individual resonances are identified in \textbf{b} and explained in the main text. Most relevant is the resonance $\vert D \rangle$, which corresponds to a two-qubit hybrid state with an anti-symmetric superposition of the two ALQ excited states. The increase contrast for features at high $f_\mathrm{drive}$ is caused by the vicinty to the symmetric coupler mode frequency. \textbf{b,} Calculated excitation spectrum as a function of the $V_\mathrm{R}$ using the Tavis-Cummings model with two qubits, one resonator mode and the parameters found indepdently in Fig.\,\ref{fig:fig2}. The plotted spectrum contains excitation spectra of different parity configurations distinguished by colors ($\mathcal{E}$ = single ALQ even parity, $\mathcal{O}$ = single ALQ odd parity). The dashed black lines correspond to the two hybrid states with a symmetric superposition of the two qubit states while the solid line corresponds to that with an anti-symmetric superposition of the two qubit states.}
 \label{fig:fig3}
\end{figure*}

The main result of our work is shown in Fig.\,\ref{fig:fig3}). We investigate the circuit with both qubit frequencies tuned into resonance with the anti-symmetric coupler mode. First, the left ALQ-coupler maximally hybridized states $ \vert \pm_\mathrm{L}\rangle$ is generated by setting $V_\mathrm{L}=1.810\,\mathrm{V}$ and $\delta_\mathrm{L}=\pi$. Then, we perform again pulsed two-tone spectroscopy while sweeping $V_\mathrm{R}$, such that the qubit frequency $f_\mathrm{qb,R}$ of the R-ALQ crosses $f_\mathrm{a}$. The differential quadrature $\Delta Q$ between the two probe pulses is plotted as a function of $V_\mathrm{R}$ and $f_\mathrm{drive}$ in Fig.\,\ref{fig:fig3}a). The measurement exhibits a spectrum showing that a resonance evolves from $ \vert -_\mathrm{L}\rangle$ for $f_\mathrm{qb,R}\ll f_\mathrm{a}$ to $ \vert +_\mathrm{L}\rangle$ for $f_\mathrm{qb,R}\gg f_\mathrm{a}$. As the single qubit frequency $f_\mathrm{qb,R}$ is tuned into resonance with $ \vert -_\mathrm{L}\rangle$, the R-ALQ hybridises with the left-site coupled system, forming an avoided crossing between the resonances of $ \vert -_\mathrm{L}\rangle$ and the R-ALQ excited state. Likewise, a second avoided crossing emerges when $f_\mathrm{qb,R}$ crosses with the transition frequency to $ \vert +_\mathrm{L}\rangle$, giving rise to the sigmoid-like dispersion \cite{PhysRevLett.103.083601,PhysRevX.8.041018,Borjans2020}. This resonance is characteristic for the two-qubit hybrid state $\vert D \rangle$ with an anti-symmetric superposition of the two qubit states and demonstrates a strong coupling between the two ALQs over a macroscopic distance. 

To identify the eigenstates of the complete circuit, we calculate the excitation spectrum as a function of $V_\mathrm{R}$ using the Tavis-Cummings model $H_\mathrm{TC} = \sum^N_\mathrm{i} \hbar g_\mathrm{a,i} (a\sigma_\mathrm{+,i}+a^\dagger\sigma_\mathrm{-,i})$ \cite{PhysRev.170.379}. It describes the interaction between one resonator mode and $N > 1$ qubits. Inserting the qubit-coupler coupling rates extracted from single qubit measurements in Fig.\,\ref{fig:fig2}), the resulting dispersion relation of the eigenstates reproduces the experimental data very well (Fig.\,\ref{fig:fig3}b). The solid black line indicates the transition to the excited state $\vert D \rangle$. This model now allows us to identify the eigenstates, where the sigmoid-like resonance is a superposition of the two bare qubit excited states and the resonator state. When all three bare transition frequencies are resonant (black arrows), the coherent exchange between the qubits result in an eigenstate $\vert D_0 \rangle$ devoid of the photonic excitation. This state explicitly reads
\begin{equation}
\vert D_0 \rangle = \frac{1}{\sqrt{g_\mathrm{c,L}^2+g_\mathrm{c,R}^2}}\big ( g_\mathrm{c,R}\vert 0,e_\mathrm{L},g_\mathrm{R} \rangle-g_\mathrm{c,L}\vert 0,g_\mathrm{L},e_\mathrm{R}\rangle \big ).
\end{equation}
The dashed black lines are associated with transitions to the two two-qubit hybrid states $\vert \pm \rangle$ with a symmetric superposition of the two ALQ excited states. The existence of these strongly correlated quantum states formed by the exchange of a cavity photon shows that both ALQs, in the even parity $\mathcal{EE}$, coherently interact with an exchange-type coupling and is the main result of this work. However, the spectrum is more complex and requires an additional consideration of parity switching events.

Because parity switching events are expected to occur locally and uncorrelated on the two ALQs, four parity configurations can randomly happen during the measurement, namely a single ALQ in an odd parity and the other in the even parity, denoted $\mathcal{OE}$ and $ \mathcal{EO}$, or both ALQs in the odd or even parity, denoted $\mathcal{OO}$ and $\mathcal{EE}$, respectively. The eigenenergies of the coupled system in the $\mathcal{OE}$, $\mathcal{EO}$ and $\mathcal{OO}$ configurations are all shown in Fig.\,\ref{fig:fig3}b). The parity configuration $\mathcal{EO}$ gives rise to two constant resonances that are associated with the transition to the states $ \vert \pm_\mathrm{L}\rangle$, reproducing the measurement Fig.\,\ref{fig:fig2}b, because the R-ALQ in the odd parity is not coupled to the resonator. The parity configuration $\mathcal{OE}$ adds an avoided crossing between the bare anti-symmetric mode and the R-ALQ excited state, because, here, the L-ALQ in the odd parity is decoupled from the resonator. These four resonances are seen in the experiment, showing that both ALQs switch between parities during a measurement time of $1.5\,$s. More data of the L-ALQ slightly detuned from $f_\mathrm{a}$ plotted in the Extended data Fig.\,\ref{fig:ext4} can be reproduced in this model with the same device parameters, showing that this quantum system is very versatile, well controlled and well-understood.

In summary, we demonstrate the remote coupling of two ALQs over a distance of six millimeters, mediated by a specifically designed superconducting resonator with a strongly and a weakly coupled mode for qubit readout and qubit coupling, respectively. In the resonant regime, the corresponding eigenstates of the full hybrid circuit can be identified using independently determined parameters inserted into the Tavis-Cummings Hamiltonian with two qubits and one cavity mode. At the point at which all energies are degenerate, we find a maximally entangled two-qubit state, mediated by the cavity coupler. Our experiments form a proof-of principle for cavity-mediated interactions between ALQs which might be potentially useful for complex quantum computer architectures. In the future, our results could be transferred to remote Andreev spin qubits in which the spin of an Andreev level is exploited as a qubit.

\section*{Bibliography}
\bibliography{sn-bibliography}
\clearpage

\section*{Methods}
\textbf{Fabrication.} Our device is fabricated on an undoped Si substrate with the native SiO$_2$ removed by a wet etching process using HF and Piranha \cite{ungerer2023performance}. We design and optimize the microwave circuit using Sonnet Suites$^\mathrm{TM}$, and then we pattern the cavity coupler and control lines by direct laser writing and reactive ion etching of a sputtered NbTiN thin film. The
latter has a thickness of 85\,nm with a sheet kinetic inductance of 2.4\,pH/$\square$. Ti/Pd (5\,nm/25\,nm) metallic gate structures are deposited by e-beam lithography and e-beam evaporation. Next, wurtzite
InAs nanowires with a 30\,nm thick epitaxial Al full-shell are suspended across the gap in the NbTiN layer using a micromanipulator. The segment of Al shell suspended over the gate structures is then removed by a 18\,s long etch step in Transene type D. The left junction is 190$\,$nm long and the right junction is 280$\,$nm. The nanowires are contacted to the rest of the circuit by 125\,nm thick Al after 27\,s of in-situ Ar-milling. The Al evaporation is performed at a substrate temperature of -20°C to reduce granularity of the Al film. Because the loop inductance is much smaller than the junction inductance, the phase difference induced by the magnetic flux mostly drops across the junction, such that $\Phi_{\rm L,R}=\Phi_0\delta_\mathrm{L,R}/2\pi$, with $\Phi_0=\frac{h}{2e}$ being the superconducting flux quantum.

\noindent\textbf{Capacitively coupled quarter-wavelength transmission line resonators.} The cavity coupler is composed of two quarter-wavelength coplanar transmission line resonators coupled by an interdigitated capacitor with a designed capacitance of $43\,\mathrm{pF}$. The resonator symmetry refers to the symmetry of the voltage amplitude. In addition to the symmetric half-wavelength coupler mode, the coupling capacitor allows an anti-symmetric half-wavelength mode, whose voltage has a phase difference of $\pi$ across the capacitor. The coupling rate of the coupler into the readout port depends on the design of the port. In our design with a single-ended capacitively coupled readout port, the symmetric resonator fields with zero phase difference across the capacitor couple to the measurement circuit with a finite coupling rate $\kappa_\mathrm{s}$ while the coupling rate of the anti-symmetric mode $\kappa_\mathrm{a}$ is considerably smaller [Extended data Fig.\,\ref{fig:ext1}].

\noindent\textbf{Pulsed two-tone spectroscopy}. We apply two-stage microwave pulses to the readout port for qubit excitation and coupler probe [Extended data Fig.\,\ref{fig:ext2}]. In the excitation stage, a strong excitation pulse and a weak coupler probe pulse are switched on simultaneously for 8$\,\mathrm{\mu s}$. The microwave frequency of the excitation pulse $f_\mathrm{drive}$ is varied to obtain the measured spectra in the main text. The coupler probe pulse is composed of four $2\,\mathrm{\mu s}$-pulses with a microwave frequency at $\sim6.461\,\mathrm{GHz}$. The pulse length of $2\,\mathrm{\mu s}$ for a single readout pulse is imposed by the hardware. It is reflected from the coupler and travel through a dimer Josephson-junction-array amplifier \cite{PhysRevApplied.13.024015}, a HEMT, and a room-temperature
amplifier before the data acquisition. In the relaxation stage, microwave pulses are switched off for 8$\,\mathrm{\mu s}$ such that the system can relax to the ground state. A weak coupler probe pulse is subsequently switched on, reflected and acquired to probe the hybrid circuit's ground state. Finally, the differential quadrature $\Delta Q$ between the two reflected probe pulses is computed and plotted as spectra in the main text.

\noindent\textbf{Calculation of the excitation spectrum of the Tavis-Cummings model}. For the single ALQ-coupler system discussed in Fig.\,\ref{fig:fig2}c), we use the Hamiltonian
\begin{equation}
    H_\mathrm{N=1} = h f_\mathrm{a} a^\dagger a + \frac{h f_\mathrm{qb,R}}{2}\sigma_{z,\mathrm{R}}+ \hbar g_\mathrm{a,R} (a\sigma_\mathrm{+,R}+a^\dagger\sigma_\mathrm{-,R}),
\end{equation}
with $\sigma_{z,\mathrm{R}}$ being the qubit Pauli z-operator for the R-ALQ. The anti-symmetric mode resonance frequency $f_\mathrm{a}$ and the qubit-coupler coupling strength $g_\mathrm{a,R}$ are extracted from Fig.\,\ref{fig:fig2}b). The gate voltage dependence of the qubit frequency $f_\mathrm{qb,R}(V_\mathrm{R})$ is obtained by linearly interpolating the measured transition frequencies to $\vert\pm_\mathrm{R}\rangle$ at the far-resonant points, at which the hybrid state mainly consists of the qubit excited state. The Hamiltonian is numerically diagonalized to obtain its eigenenergies. To account for the decoupled odd parity, the eigenenergies are computed with $g_\mathrm{a,R}=0$ and the spectrum is overlaid on top of that of the even parity eigenenergies.

For the full hybrid circuit discussed in Fig.\,\ref{fig:fig2}d), the Hamiltonian
\begin{equation}
    H_\mathrm{N=2} = h f_\mathrm{a} a^\dagger a + \sum_\mathrm{i \in {L,R}} \bigg [ \frac{h f_\mathrm{qb,i}}{2}\sigma_{z,\mathrm{i}}+ \hbar g_\mathrm{a,i} (a\sigma_\mathrm{+,i}+a^\dagger\sigma_\mathrm{-,i}) \bigg ]
\end{equation}
is diagonalized separately for four set of parameters to compute the excitation spectra in the four different parity configurations. The four spectra are then overlaid to obtain Fig.\,\ref{fig:fig2}d). The corresponding coupling strengths are listed in the Tab.\,\ref{tab1}.
\begin{table}[h]
\caption{Coupling strengths used to compute excitation spectra in different parity configurations.}\label{tab1}%
\begin{tabular}{@{}lll@{}}
\toprule
Parity configurations & $g_\mathrm{a,L}/(2\pi) \: (\mathrm{MHz})$  & $g_\mathrm{a,R}/(2\pi) \:  (\mathrm{MHz})$\\
\midrule
$\mathcal{EE}$  & 141   & 104\\
$\mathcal{EO}$  & 141   & 0\\
$\mathcal{OE}$  & 0   & 104\\
$\mathcal{OO}$  & 0   & 0\\
\botrule
\end{tabular}
\end{table}

\section*{Acknowledgements}
We thank Peter Krogstrup for assistance on material growth. We thank Joost Ridderbos, Marcelo Goffman and Hugues Pothier for assistance on the microwave measurement. We thank Alfredo Levy Yeyati and  Francisco Matute-Cañadas for scientific discussions about the physics of Andreev bound states. We thank Nicolas Zapata for assistance on the fabrication and characterization of the JPA. This research was supported by the European Union’s Horizon 2020 research and innovation program through the FET-open project AndQC, agreement No 828948 and the Marie Skłodowska Curie COFUND project QUSTEC, agreement No 847471. It was further supported by the Swiss National Science Foundation through grant 192027, the NCCR Spin Qubit in Silicon (NCCR-Spin) and the Swiss Nanoscience Institute (SNI). T.K. and J.N. acknowledge financial support from the Novo Nordisk Foundation SolidQ project and Danish National Research Foundation (DNRF 101).

\section*{Author contributions statement}
L.Y.C. fabricated the quantum circuitry with the help from R.H., A.K. and J.H.U. C.C. fabricated the parametric amplifier with the assistance of P.W., T.R. and I.M.P. L.Y.C., R.H. and A.K. performed the measurements. L.Y.C. and A.K. analyzed the data with inputs from R.H., A.B. and C.S. L.Y.C. wrote the manuscript with inputs from all authors. C.S. initiated the project. T.K. and J.N. developed nanowire materials.

\section*{Competing interests statement}
The authors declare no competing interests.

\section*{Data availability statement}
All data in this publication are available in numerical form at: 
\url{https://doi.org/10.5281/zenodo.10037257}

\section{Extended data}
\renewcommand\thefigure{E\arabic{figure}}   
\setcounter{figure}{0}
\begin{figure*}[htbp]
	\centering
	\includegraphics[width=\linewidth]{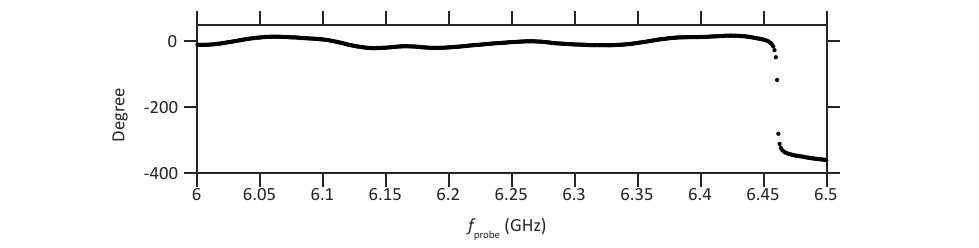}
	\caption{\textbf{Coupler spectrum.} The phase of a weak microwave probe pulse as a function of the probe frequency $f_\mathrm{probe}$. No phase shift is observed at $6.075\,\mathrm{GHz}$, showing that the anti-symmetric coupler mode is weakly coupled to the readout port with considerably smaller $\kappa_\mathrm{a}$. At $6.461\,\mathrm{GHz}$, a phase shift of 360° is measured, which results from the reflection of the symmetric coupler mode.}
 \label{fig:ext1}
\end{figure*}
\begin{figure*}
	\centering
	\includegraphics[width=\linewidth]{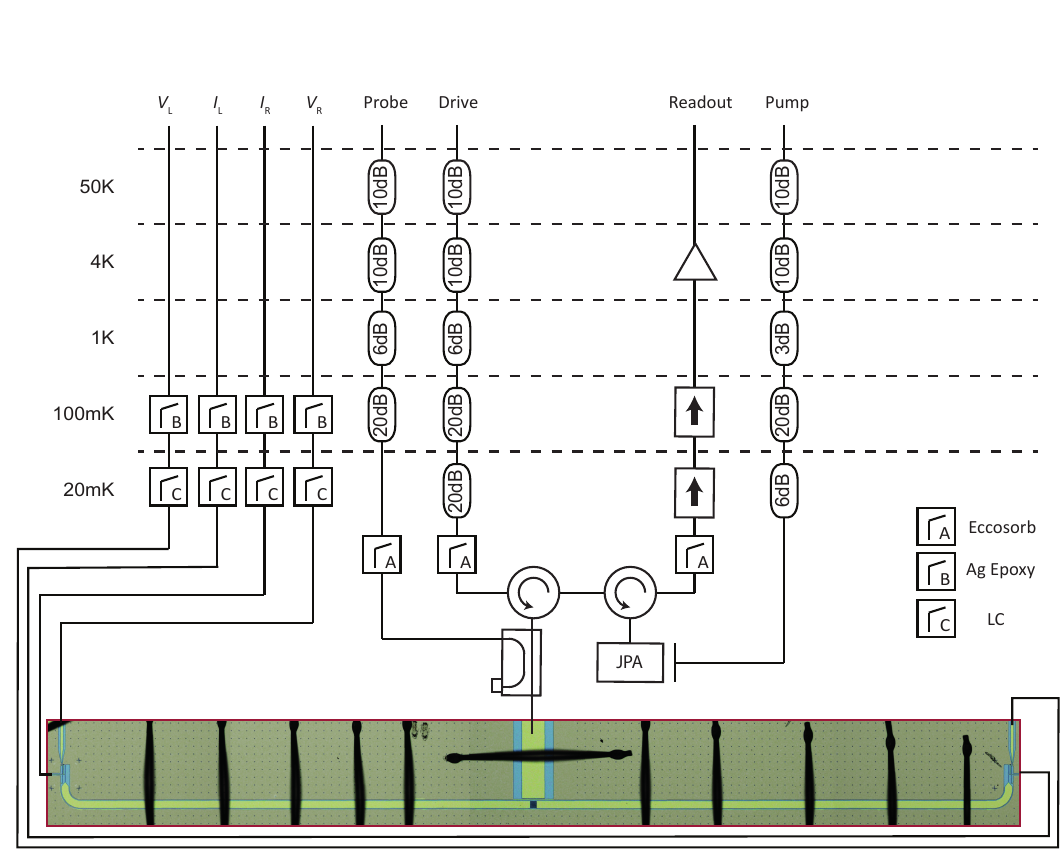}
	\caption{\textbf{Cryogenic wiring diagram.} The qubit excitation and coupler probe pulses pass through the illustrated setup before being routed to the readout port on the chip. The probe pulse is reflected and then routed through the amplification chain containing a dimer Josephson-junction-array amplifier (JPA), a HEMT, and a room-temperature amplifier. An additional RF line is used to pump the JPA. The pump, drive and probe RF lines are filtered with eccosorb at the mixing chamber. The DC voltage and current lines are routed to the chip through Ag epoxy at the coldplate for thermal anchoring and three-stage LC filters with lowest cutoff frequencies at 80\,MHz at the mixing chamber.}
 \label{fig:ext2}
\end{figure*}
\begin{figure*}
	\centering
	\includegraphics[width=\linewidth]{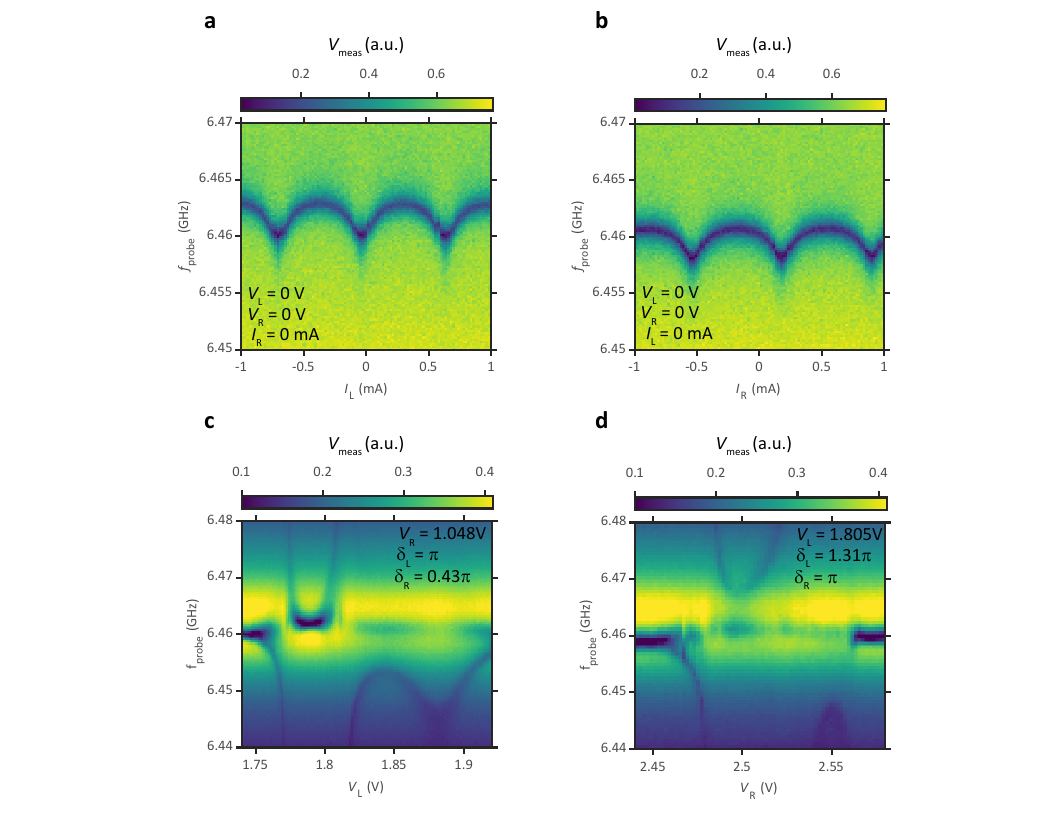}
	\caption{\textbf{Reflection spectrum of the symmetric coupler mode.} \textbf{a,} The voltage of a reflected weak coupler probe pulse as a function of the probe frequency $f_\mathrm{probe}$ and DC current $I_\mathrm{L}$ near the symmetric mode frequency $f_\mathrm{s}$. The symmetric mode resonance is periodically modulated by flux biasing the left rf-SQUID. Similar behavior is observed when changing $I_\mathrm{R}$, as depicted in \textbf{b}. The $\pi$-points are identified as the current values with the lowest resonance frequency. \textbf{c-d,} A transition of the L-ALQ (R-ALQ) is tuned across $f_\mathrm{s}$ at $\delta_\mathrm{L}=\pi$ ($\delta_\mathrm{R}=\pi$), creating a gate dependent shift of the symmetric mode resonance. The resonance at $f_\mathrm{s}$ in addition to the shifted resonance is characteristic for Andreev qubits and indicates finite dwell time in the odd parity. The gate voltage independent background is caused by the finite bandwidth of the JPA.}
 \label{fig:ext3}
\end{figure*}
\begin{figure*}
	\centering
	\includegraphics[width=\linewidth]{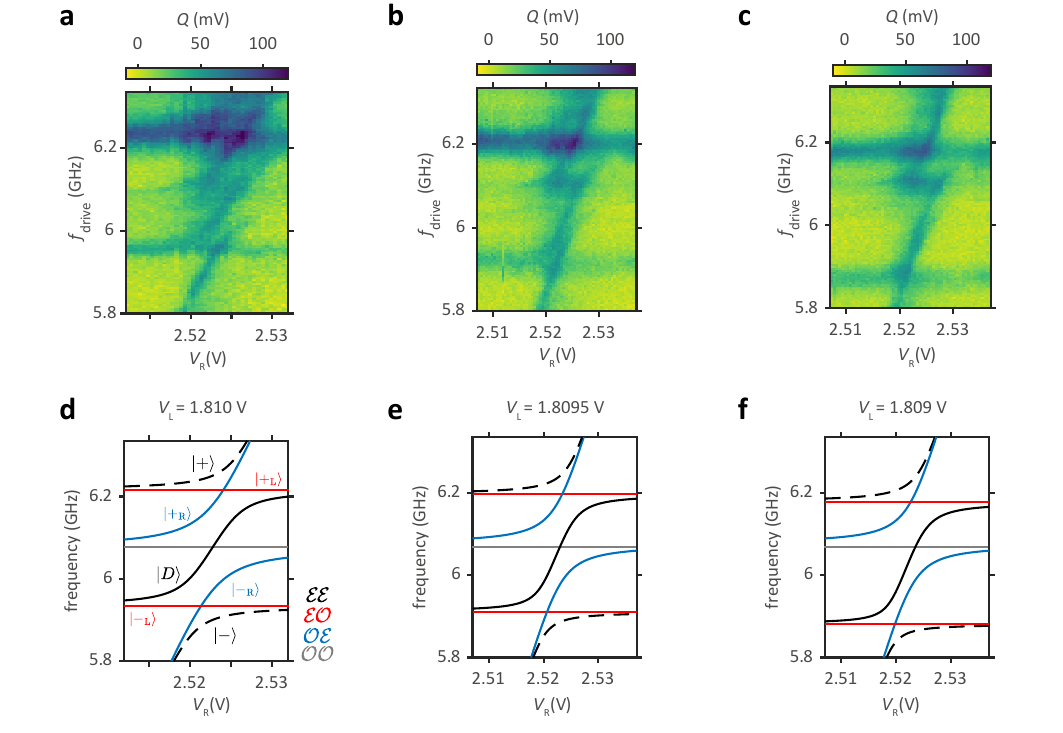}
	\caption{\textbf{Extended spectroscopy measurements of the full hybrid circuit at different $V_\mathrm{L}$.} \textbf{a} and \textbf{d} are also shown in the Fig.\,\ref{fig:fig3} in the main text. The spectra in \textbf{b} and \textbf{c} are measured by step-wise reducing $V_\mathrm{L}$. \textbf{e} and \textbf{f} are calculated spectra using the Tavis-Cummings model with $f_\mathrm{qb,L}=0.993\,f_\mathrm{a}$ and $f_\mathrm{qb,L}=0.985\,f_\mathrm{a}$. Crucially, as the L-ALQ is detuned from the anti-symmetric mode frequency $f_\mathrm{a}$, the sigmoid-shaped two-qubit entangled state $\vert D \rangle$ (solid black) approaches the single ALQ-coupler state $\vert -_\mathrm{R} \rangle$ (blue) from the parity configuration $\mathcal{OE}$, in agreement to the measured spectra in \textbf{b} and \textbf{c}.}
 \label{fig:ext4}
\end{figure*}
\end{document}